\documentclass{jfm}
\usepackage{amsmath}
\usepackage{amsfonts}
\usepackage{amssymb}
\usepackage{amsbsy}
\usepackage{graphicx}
\usepackage{natbib}
\usepackage{bm}

\newcommand\bsigma{\bm{\sigma}}
\newcommand\Qmat{\ensuremath{\mathsfbi{Q}}}
\newcommand\Bmat{\ensuremath{\mathsfbi{B}}}
\newcommand\Imat{\ensuremath{\mathsfbi{I}}}
\newcommand\Cmat{\ensuremath{\mathsfbi{C}}}

\newcommand{\AVE}[1]{\langle{#1}\rangle}
\newcommand{\te}[1]{\bm{\ensuremath{ #1 }}}

\newcommand\bx{\te{x}}

\newcommand\bU{\te{U}}

\newcommand\bj{\te{j}}

\newcommand\bF{\te{F}}

\newcommand\bn{\te{n}}

\newcommand\bq{\te{q}}

\newcommand\bbm{\te{m}}

\newcommand\balpha{\te{\alpha}}

\newcommand\dup{\ensuremath{\,\mathup{d}}}
\newcommand\Piup{\ensuremath{\mathup{\Pi}}}
\title{The force on a boundary in active matter}

\author[Wen Yan and John F. Brady]%
{Wen Yan$^1$\break
	and John F. Brady$^2$\thanks{Email address for correspondence: jfbrady@caltech.edu}}

\affiliation{$^1$Department of Mechanical \& Civil Engineering, California Institute of Technology,
	Pasadena, CA 91106, USA\\[\affilskip]
	$^2$Divisions of Chemistry \& Chemical Engineering and Engineering \& Applied Science, California Institute of Technology,
Pasadena, CA 91106, USA}

\date{?; revised ?; accepted ?. - To be entered by editorial office}

\begin{document}

\maketitle

\begin{abstract}
We present a general theory for determining the force (and torque) exerted on a boundary (or body) in active matter.  The theory extends the description of passive Brownian colloids to self-propelled active particles and applies for all ratios of the thermal energy $k_BT$ to the swimmer's activity $k_sT_s = \zeta U_0^2\tau_R/6$, where $\zeta$ is the Stokes drag coefficient, $U_0$ is the swim speed and $\tau_R$ is the reorientation time of the active particles.  The theory %,  which is valid on all length and time scales, 
has a natural microscopic length scale over which concentration and orientation distributions are confined near boundaries, but the microscopic length does not appear in the force.  The swim pressure emerges naturally and dominates the behavior when the  boundary size is large compared to the swimmer's run length $\ell  = U_0 \tau_R$.   The theory is  used to  predict the motion of bodies of all sizes immersed in active matter.
\end{abstract}


%example of including figure
%\begin{figure}
%	\centerline{\includegraphics{trapped.eps}}% Images in 100% size
%	\caption{Trapped-mode wavenumbers, $kd$, plotted against $a/d$ for
%		three ellipses:\protect\\
%		---$\!$---,
%		$b/a=1$; $\cdots$\,$\cdots$, $b/a=1.5$.}
%	\label{fig:ka}
%\end{figure}
\section{Introduction}
The behavior of self-propelled objects such as bacteria, algae, and  synthetic Janus particles has become an dynamic field of research,  both for the `swimming' of individual particles  
\citep{HydroSwimming2009},  
and for the collective behavior of active suspensions \citep{Toner05}.  Owing to the particles' self motion, active matter can  spontaneously phase separate into dense and dilute regions \citep{Cates10, Bialke13, Buttinoni13, Stenhammar13, Palacci13, Wysocki14, Fily12, Pressure2014, Marchetti2104, Stenhammer14, ThermoActive2015} and  can move collectively under an orienting field \citep{PolarField2014}.  
	
	Recently, the swim pressure~\citep{Pressure2014, Marchetti2104} was introduced as a new perspective 
	on the behavior of active matter.  The swim pressure is the pressure needed to confine active particles and is analogous to the osmotic pressure of Brownian colloids. %For example, active Brownian particles (ABPs) that  separate into dilute and dense regions are now understood as a `gas-liquid' coexistence with the swim pressure providing the necessary link.  %The decrease in the swim pressure with  concentration destabilizes the system resulting in phase separation~\cite{ThermoAct2015}.   The swim pressure is the pressure needed to confine the active particles.  
	%The swim pressure is the force per unit area exerted by active particles on the confining walls 
	%and is thus intimately linked to the particle diffusivity.  
	The dilute limit  `ideal gas' swim pressure $\Piup^{swim} = n \zeta U_0^2\tau_R/6$ (in 3D), where $n$ is the number density of active particles, $\zeta$ is their drag coefficient, $U_0$ is the swim speed, and $\tau_R$ is their reorientation time.    
	The swim pressure, or stress, is  defined as  the moment of the swim force $\AVE{\bsigma^{swim}} =-n\AVE{\bx\bF^{swim}}$, where  $\bF^{swim}=\zeta U_0 \bq$, with $\bq$ the orientation vector of the swimmer and $\bx$ its  position.  %The position is simply $\bx(t) = \int^t U_0 \bq(t^\prime)dt^\prime$, and thus, $\bsigma^{swim}  = - n \zeta U_0^2 \int^t\langle \bq(t)\bq(t^\prime)\rangle dt^\prime = - n \zeta U_0^2 \tau_R/6\, \Imat$ (for times $t \gg \tau_R$), arising from the random reorientation of the swimmer. % $\langle \bq(t)\bq(t^\prime)\rangle  = (\Imat /3) \exp\{{-2(t-t^\prime)/\tau_R}\}$. 
	The `moment arm' for the swim stress is the swimmer's run length,  $\ell =  U_0\tau_R$. 
	%, and hence a swim stress $\bsigma^{swim} - n \zeta U_0^2 \tau_R/6\, \Imat$. % (for times $t \gg \tau_R$).
	
	%and thus, $\bsigma^{swim}  = - n \zeta U_0^2 \int^t\langle \bq(t)\bq(t^\prime)\rangle dt^\prime = - n \zeta U_0^2 \tau_R/6\, \Imat$ (for times $t \gg \tau_R$), arising from the random reorientation of the swimmer: $\langle \bq(t)\bq(t^\prime)\rangle  = (\Imat /6) \exp\{{- (t-t^\prime)/\tau_R}\}$.
	
	%Important in the definition of 
	The swim pressure is an average over the reorientation time $\tau_R$, which implies an average over the run length $\ell$.  The swim pressure is only defined on, and applies for, lengths greater than the run length.  And its use to compute forces on boundaries necessitates that the boundary or macroscopic length scale, $L$, be much larger than the run length \citep{Yan2015}.  What happens when the length scale of interest is not large compared to the run length?  %What description is now appropriate?  
	Can  we extend the notion of the swim pressure to such situations?  Or more generally, how does the swim pressure emerge from a more microscopic description?
	
	In this paper we provide such a microscopic theory and show how the swim pressure arrises naturally. % as the characteristic length scale becomes large compared to the run length.  
	The theory is an extension of the well-known dynamics of passive colloidal particles to active colloidal particles, and will allow us to compute forces and torques on bodies and thus  predict their motion in response to the swimmers' activity.
	
	\section{Theory}
	
	For active colloidal particles there are three characteristic lengths:  (i) the macroscopic length scale $L$, (ii) the run length $\ell = U_0\tau_R$ and (iii) a microscopic length $\delta = \sqrt{D_T \tau_R}$, where $D_T$ is the translational diffusivity of the active particles.  %In the common situation of ABPs the product of the translational Brownian diffusivity and the reorientation time gives $\delta \sim a$, where $a$ is the size of the active swimmer.  
	Although in a typical application we expect  $L > \ell \gg \delta$, the theory we  present is  valid for any ratio of length scales. %, even if the `body' is small compared to the swimmer's size.
	
	%\subsection{A Microscopic Theory of Active Matter}
	%\label{sec:theory}
	%Active Brownian particles (ABP) are governed by the overdamped Langevin equation
	%\begin{equation}\label{eq:ABP}
	%0 = - \zeta\bU +  \bF^{swim} + \bF^{P} + \bF^B  \, ,
	%\end{equation}
	%where $\bU$ is the particle velocity, $\bF^P$ is a boundary force, and $\bF^B$  is the Brownian force responsible for the translational diffusivity $D_T = k_BT/\zeta$.  The orientation vector $\bq$ in the swim force is subject to rotational Brownian diffusion ($D_R = 1/\tau_R$), and follows directly from a torque  balance. For a spherical swimmer, $\zeta=6\upi\eta a$, where $a$ is the swimmer radius and $\eta$ is  the viscosity of the suspending Newtonian fluid.  For reorientation due to Brownian rotation, $D_R = k_BT/8\upi \eta a^3$, and the microscopic length scale $\delta = \sqrt{D_T/D_R} = \sqrt{4/3}a$.
	
	Active Brownian particles (ABP) are governed by the Smoluchowski equation for the probability density for finding a swimmer at $\bx$ with orientation $\bq$:
	\begin{equation}
	\frac{\partial P(\bx,\bq,t)}{\partial t} + \nabla \cdot \bj^T + \nabla_{R} \cdot \bj^R = 0 \, .
	\label{eq:Smol}
	\end{equation}
	The translational and rotational  fluxes are:
	%\begin{eqnarray}
	$ \bj^T =   (U_0 \bq + \bF^P/\zeta - D_T \nabla \ln P)P$, and 
	$ \bj^R = - D_R \nabla_{R} P$ ,
	% \label{eq:fluxes}
	% \end{eqnarray}
	where $\nabla_R = \bq\times \nabla_{\bq}$ is  the orientational gradient operator.   For a spherical swimmer  of radius $a$, $\zeta=6\upi\eta a$, $D_T = k_BT/6\upi\eta a$,  $D_R ( =1/\tau_R) = k_BT/8\upi \eta a^3$ and the microscopic length scale $\delta = \sqrt{D_T/D_R} = \sqrt{4/3}a$.
	
	%The orientation vector $\bq$ in the swim force is subject to rotational Brownian diffusion ($D_R = 1/\tau_R$), and follows directly from a torque  balance. For a spherical swimmer, $\zeta=6\upi\eta a$, where $a$ is the swimmer radius and $\eta$ is  the viscosity of the suspending Newtonian fluid.  For reorientation due to Brownian rotation, $D_R = k_BT/8\upi \eta a^3$, and the microscopic length scale $\delta = \sqrt{D_T/D_R} = \sqrt{4/3}a$.
	
	At a boundary surface the normal component of the translational flux must vanish,  $\bn\cdot\bj^T =0$.  If there were no translational Brownian motion or boundary force, then $U_0(\bn \cdot \bq) P =0 $, which means that either (i) $U_0 = 0$ or (ii) $\bn\cdot\bq = 0$ or (iii)  $P=0$ at the surface; none of which is true in general.  It is essential to have a strong enough boundary force ($\bF^P$) or translational Brownian diffusion (or both, or hydrodynamics) to prevent particle crossing.  As is well known in colloidal dynamics, a hard-particle repulsive force is infinite and nonzero only at the boundary surface and the no flux condition is satisfied via the Brownian flux. %: $U_0\bn \cdot \bq P  -D_T \bn\cdot \nabla P = 0$.  
	
	Rather than use a finite range and amplitude boundary force or hydrodynamic lubrication interactions  %that would make $\zeta$ a function of the distance from the surface 
	to prevent particle flux, we choose to use $D_T$ as this is the simplest to implement theoretically and most easily reveals the underlying physics. It is important to note that whatever means is used to prevent active particles from crossing a boundary it will introduce a microscopic length scale $\delta$.   As we shall see, for pressures and forces, $\delta$ will not appear in the final results.  Any of the mechanisms would produce the same behavior.  
	
	Indeed, \cite{Ezhilan2015} recently examined active particles in 2D confined between two walls without translational Brownian motion  ($D_T \equiv 0$)  and showed that the problem could be modeled with two regions:  freely swimming bulk behavior connected to a singular surface layer of particles in contact with the walls.  The action of translational Brownian motion is to spread this singular surface layer over the microscopic  thickness $\delta$ adjacent to the walls, as is standard in boundary-layer theory.  Our planner 2D results are in agreement with their findings.

	% Although we speak in terms of translational Brownian motion and forces proportional to $k_BT$, this is not necessary.  One can simply replace $k_BT$ with $\zeta D_T$ and the results are unchanged; the translational diffusion, like the rotary diffusion $D_R$, need not be thermal in origin. 
	
	The Smoluchowski equation applies for all length and time scales but its solution in any but the simplest situations is challenging. %~\footnote{In 2D there are two spatial  and one angular coordinate, while in 3D there are now six coordinates, precluding any general solution.}.  
	We need a simplified description that captures the essential physics, and, more importantly, provides insight into the general behavior and can explain phenomena without detailed calculations.  Consider a body immersed in a dilute suspension of ABPs.  With  $\bF^P =0$, the force the active colloidal particles exert on the body is given exactly by \citep{Brady1993, Squires2005}
	%\begin{equation}
	$ \bF = - k_BT \int_{S_B} P(\bx,\bq,t) \bn dS \, ,
	\label{eq:forceonbody}$
	% \end{equation}
	where $\bn$ is the outer normal to the body surface as shown in Fig.~\ref{fig:asym}.  The force averaged over the orientation of the active particles is
	\begin{equation}
	\AVE{ \bF}_{\bq} = - k_BT \int_{S_B} n(\bx,t) \bn \mathrm{d} S \, ,
	\label{eq:forceonbody2}
	\end{equation}
	where $n(\bx,t) \equiv \int P(\bx,\bq,t) d\bq$ is  the number density of swimmers.% which requires a conservation equation for $n(\bx,t)$. % rather than $P(\bx,\bq,t)$.
	
	The conservation equations for the zeroth and first moments of the Smoluchowski equation are~\citep{Saintillan2015}:
	\begin{eqnarray}
	\frac{\partial n}{\partial t} & + &  \nabla\cdot \bj_n = 0\ \   ,\  \  \bj_n  =  U_0\bbm - D_T \nabla n   \, , \label{eq:jn} \\*
	\frac{\partial \bbm}{\partial t} & + &  \nabla\cdot  \bj_m + 2D_R \bbm = 0\, ,  \label{eq:jm} \\*
	\mbox{with} \quad \bj_m & =&   U_0\Qmat  +  \textstyle{\frac{1}{3}}U_0 n\, \Imat - D_T \nabla\bbm \,  ,
	\label{eq:Q}
	\end{eqnarray}
	where  $\bbm(\bx,t) = \int \bq P(\bx,\bq,t) \mathrm{d}\bq$ is  the polar order field,  and $\Qmat(\bx,t) = \int (\bq\bq- \frac{1}{3}\Imat) P(\bx,\bq,t)\mathrm{d}\bq$ is the nematic order  field.  Since the force on a body only involves the number density at the surface, we can use the simplest closure of the hierarchy $\Qmat  = 0$.  We show below (and discuss in  Appendix B) that this closure is sufficient to achieve good accuracy and reveals the essential physics.
	
	%with  fluxes 
	%\begin{eqnarray}
	%\bj_n & =&  U_0\bbm - D_T \nabla n \, , \\*
	%\bj_m & =&   U_0\bQ  +  \textstyle{\frac{1}{3}}U_0 n\, \Imat - D_T \nabla\bbm \, , \\*
	%\bj_{Q} & =&  U_0\bB  - \textstyle{\frac{1}{3}} U_0   \bbm \, \Imat - D_T \nabla\bQ \, .
	%\end{eqnarray}
	%Here, $\bB = \int \bq\bq\bq P d\bq$ is the third moment and has a conservation equation analogous to (\ref{eq:Q}), and so on for higher moments.  The infinite hierarchy must be closed at some level.    % by comparing with closing at the next level -- keeping the equation for $\bQ$ with the closure on $\bB = \frac{1}{6}\alpha \odot \bbm$, where $\alpha$ is the fourth order isotropic tensor, and by comparison with direct simulations of the Langevin equation for ABPs.

	Two remarks will help understand the structure of the moment equations.  First, when departures from uniformity vary slowly, the $\bbm$-field equation has a balance between the `sink' term and the gradient in the concentration, $2D_R \bbm \approx - \frac{1}{3}U_0 \nabla n$, which gives a diffusive flux in the concentration field that incorporates the swim diffusivity: $\bj_n \approx - (D_T + \frac{1}{6} U_0^2 \tau_R) \nabla n$.  Second, at the other extreme when variations are rapid, the $\bbm$-field has a natural screening length where diffusion balances the sink: $D_T \nabla^2 \bbm \approx  2D_R \bbm$.  This screening length is proportional to the microscopic length $\delta = \sqrt{D_T/D_R}$.  The screening length  plays a fundamental, but unusual, role in active matter---it is essential in order to have a well-posed problem and there will be rapid variations in properties on the scale of $\delta$, but in the limit where $\delta \ll \ell, L$, the microscopic length does not appear in the active pressure or in the forces and torques on boundaries.  The athermal limit ($D_T \rightarrow 0$) is singular and $D_T$ can only be set to zero {\em after} the limit is taken.
	
	%As discussed earlier, the active matter problem is singular at boundaries unless there is a microscopic length scale $\delta$. 
	
	\section{Examples}
	\label{sec:examples}
	
	First,  we consider an infinite flat plate with normal along the $z$-direction; there is no macroscopic length scale. %The problem is one dimensional and there is no macroscopic length scale.   
	The $n$- and $\bbm$-fields %equations reduce to:
	%\begin{eqnarray}
	%\frac{d}{d z} \left( U_0 m_z - D_T \frac{d n}{ d z}\right) & = & 0\, .\, \\*
	%D_T \frac{d^2 m_z}{d z^2} - 2D_R m_z & = & \frac{1}{3} U_0 \frac{dn}{dz}\, ,
	%\end{eqnarray}
	are subject to no flux at $z=0$: $\bn \cdot \bj_{n,m} = 0$ and a uniform concentration and no polar order as $z \rightarrow \infty$: $n \sim  n^\infty$ and $\bbm \sim 0$.  %The transverse polar order fields are both zero, $m_x = m_y = 0$.  
	The concentration and polar order fields are simple  exponentials %decaying on the screening length
	\begin{equation}
	n = n^\infty\left(1 + \textstyle{\frac{1}{6}}(\ell/\delta)^2 e^{-\lambda z}\right)\  , \  m_z = - n^\infty \textstyle{\frac{1}{6}}(\lambda\ell) e^{-\lambda z} \, ,
	\label{eq:n(z)}
	\end{equation}
	where $\lambda = \sqrt{2(1+\frac{1} {6}(\ell/\delta)^2)}/\delta$ is the inverse screening length.  
	
	The concentration at the wall, $n(0) = n^\infty(1 + {\frac{1}{6}}(\ell/\delta)^2)$, is independent of the closure, always exceeds that far away, and can become very large as $(\ell/\delta) \rightarrow \infty$.  This `infinite' concentration applies for a {\em dilute} suspension.  It is not a build-up associated with a finite concentration of active particles.  Rather, it is the singularity alluded to earlier that results if translational Brownian motion (or a microscopic length) is not considered.\footnote{The  active particle size $a$  must be taken into account in defining the no flux surface  $z=0$.} %To detect the large build-up simulations need to examine the region of $O(\delta)$ adjacent to a surface.}.
	
	Even though the concentration can become arbitrarily large, the force per unit area or pressure on the wall from the microscopic force definition  (\ref{eq:forceonbody2}) is finite and {\em independent} of $\delta$:  $\Piup^W =  n(0) k_BT = n^\infty(k_BT + k_sT_s)$, where we have defined the swimmer's  `activity' $k_sT_s = \zeta U_0^2\tau_R/6 = k_BT\, (\ell/\delta)^2/6$.  We recognize the pressure on the wall as the {\em active pressure}---the sum of the osmotic pressure of Brownian particles plus the swim pressure.  And note that this is true regardless of the relative magnitudes of $k_BT$ and $k_sT_s$.   Also, the ratio $(\ell/\delta)^2 = 6 D^{swim}/D_T = U_0 \ell/D_T = \Pen_{\ell}$ is a P\'eclet number based on the run length measuring the relative importance of swimming to Brownian diffusion.  
	% $\frac{1}{6}(\ell/\delta)^2 = D^{swim}/D_T = \frac{1}{6}U_0 \ell/D_T = Pe_{\ell}$ 
	
	% The second  problem is active Brownian particles confined between two parallel plates separated by a distance $H$.  The concentration distribution is now
	% \begin{equation}
	% \frac{n(z)}{n_0} = 1 + \frac{1}{6}(\ell/\delta)^2\,\frac{\sinh(\lambda z) + \sinh(\lambda(H-z))}{\sinh(\lambda H)}\, ,
	%\frac{n(z)}{n_0} = 1 + \frac{1}{6}(\ell/\delta)^2\frac{(1-e^{-\lambda H})e^{\lambda z} - (1 - e^{\lambda H})e^{-\lambda z}}{e^{\lambda H} - e^{-\lambda H}}\, ,
	% \label{eq:nchannel}
	% \end{equation}
	%where the constant $n_0$ is related to the average number density of APBs in the channel $\AVE{n} = \int_0^H n(z)dz/H$.
	The second  problem is active Brownian particles confined between two parallel plates separated by a distance $H$.  The concentration distribution is $
	n(z)/n_0 = 1 + \frac{1}{6}(\ell/\delta)^2[\sinh(\lambda z) + \sinh(\lambda(H-z))]/\sinh(\lambda H)$, where the constant $n_0$ is related to the average number density of ABPs in the channel $\AVE{n} = \int_0^H n(z)\mathrm{d} z/H$.
	% \begin{equation}
	%$n_0 =   \langle n\rangle/ \left( 1 + (\ell/\delta)^2/(3\lambda H)\, (1 -  4 e^{-\lambda H} + e^{-2 \lambda H} )/(1+ e^{-2\lambda H})\right)\, .
	%  \label{eq:avgn}$
	% \end{equation}
	% \begin{equation}
	%n_0 =   \langle n\rangle \left( 1 + \frac{1}{3}\frac{(\ell/\delta)^2}{\lambda H}\frac{1 -  4 e^{-\lambda H} + e^{-2 \lambda H} }{1+ e^{-2\lambda H}}\right)^{-1}\, .
	% \label{eq:avgn}
	% \end{equation}
	The concentration is identical at both walls %, $n(0)= n(H)  = n_0(1 + \frac{1}{6}(\ell/\delta)^2)$, 
	and is the same as for a single wall with  $n_0$ replacing $n^\infty$.  In the limit of large $\lambda H$, corresponding to $\delta \ll  H$, %$n_0 \sim \langle n\rangle [1 + (\ell/\delta)^2/(3\lambda H)]^{-1}$ 
	and  when $\delta \ll \ell$, $n_0  \sim  \langle n\rangle [1 +  (\ell/H)/\sqrt{3}]^{-1}$ and the pressure at the walls becomes 
	\begin{equation}
	\Piup^W = \AVE{n}\left( k_BT + \frac{k_sT_s}{1 +  (\ell/H)/\sqrt{3}}\right) \, .
	% \Pi^W =  \frac{\langle n\rangle  (k_BT+ k_sT_s)}{1 +  (\ell/H)/(3\sqrt{3})}\, .
	\label{eq:Pchannel}
	\end{equation}
	As for a single wall the pressure is independent of the microscopic length scale $\delta$ but now depends on the ratio of the run length to the macroscopic scale $\ell/H$.  We shall see that the this behavior is generic---the influence of the run length enters as $\ell/L$.  In a simulation study \cite{Ray2014} observed that the pressure in a channel depends on the gap spacing as predicted by (\ref{eq:Pchannel}). (In 2D  the coefficient is $1/\sqrt{2}$). % \footnote{For swimmers in 2D, the coefficient is $1/\sqrt{2}$ in place of   $1/\sqrt{3}$.}.
	
	\begin{figure}%[t]
		\centering
		\includegraphics{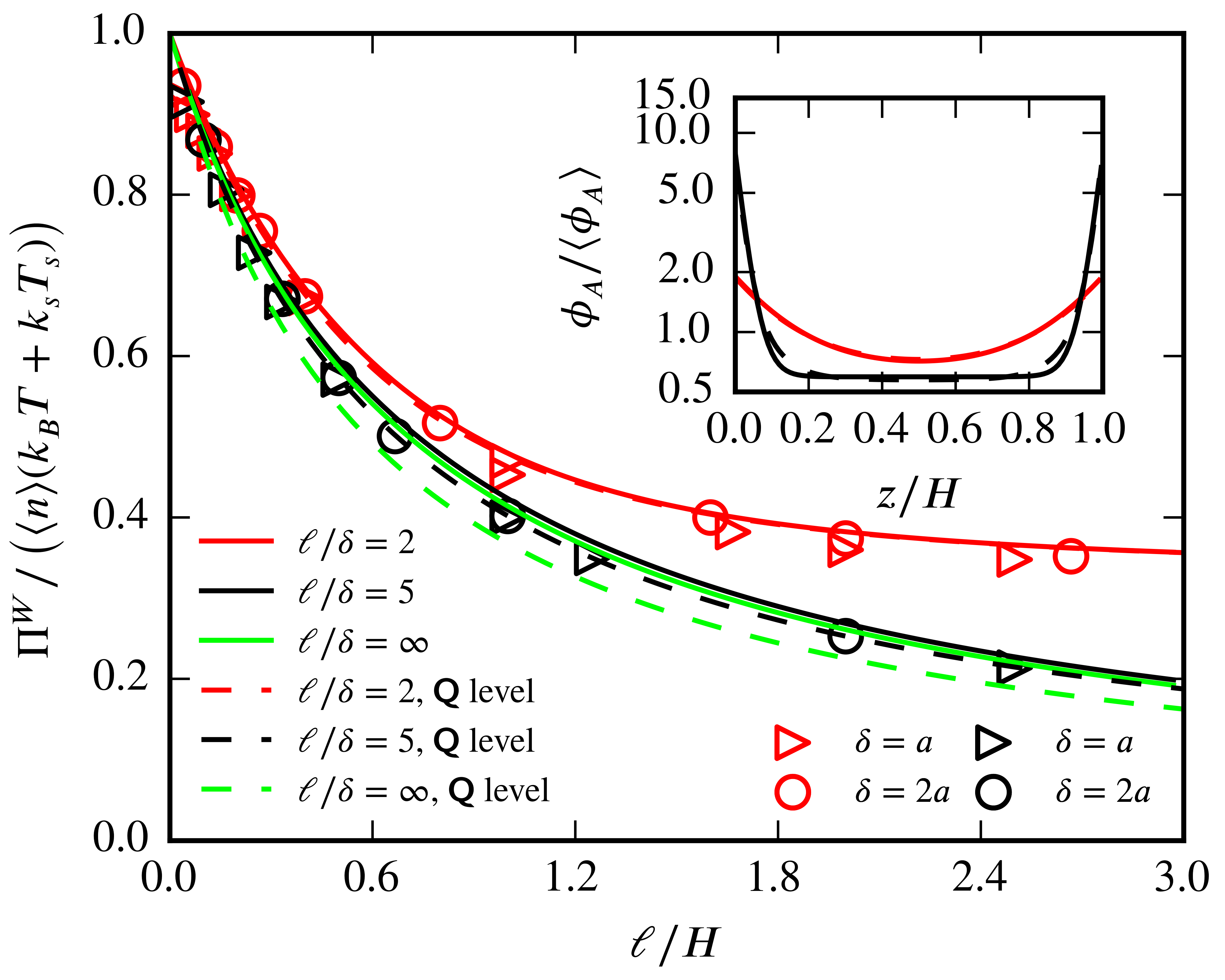}
		\caption{$\Piup^{W}$ of ABPs confined between parallel walls in 2D. The inset shows the area fraction distribution $\phi(z)$.  Here, $\ell = U_0\tau_R$ is the run length and $\delta = \sqrt{D_T\tau_R}$ is the microscopic length.  The swimmer radius is $a$.}
		\label{fig:walls}
	\end{figure}

	Figure~\ref{fig:walls} compares the concentration profile and pressure for a channel from the theory with results from ABP dynamic simulations (Appendix A).  Also shown are the theoretical predictions from closing the hierarchy at the next level including the nematic order field $\Qmat$ as described in Appendix B. % is solved with the closure $B_{ijk} = \frac{1}{4}(\delta_{ik}\delta_{jl} + \delta_{il}\delta_{jk} - \frac{2}{3}\delta_{ij}\delta_{kl} )m_l$. %(obtained from the relaxation of the $\bB$ field).  
	The $\bbm$-field closure is sufficient, both qualitatively and quantitatively.
	
	The next  problems are the concentration and pressure distribution in 3D outside and inside a sphere, and in 2D outside and inside a circle, of radius $R$.  Symmetry dictates that $n(\bx) = n^\infty f(r)$ and $\bbm(\bx) = n^\infty\bx g(r)$, where $f(r)$ and $g(r)$ are  scalar functions of $r$.  The exterior solution in 3D has the form of an exponentially screened concentration  reminiscent of Debye screening
	\begin{equation}
	\frac{n(r)}{n^\infty} = 1 + \frac{1}{6}(\ell/\delta)^2\frac{1}{1 + (1 + \lambda R)(\delta/R)^2} \frac{R}{r} e^{-\lambda(r - R)} \, ,
	\label{eq:nsphere}
	\end{equation}
	and similarly for the $\bbm$-field.  
	In 2D Bessel functions replace the exponential:
	\begin{equation}
	\frac{n(r)}{n_A^\infty} = 1 + \frac{2(\ell/\delta)^2 K_0(\lambda^\prime r)}{K_0(\lambda^\prime R)[2-(\ell/\delta)^2] + K_2(\lambda^\prime R)[2 + (\ell/\delta)^2]}\, ,
	\label{eq:ncircle}
	\end{equation}
	where the 2D inverse screening length $\lambda^\prime = \sqrt{1 + \frac{1}{2}(\ell/\delta)^2}/\delta$, $K_{0,2}$ are the modified Bessel functions and $n_A^\infty$ is the area number density at infinity.    For large $\lambda^\prime r$ the concentration disturbance  decays as $\sim e^{-\lambda^\prime r}/\sqrt{r}$.
	
	The %concentration and 
	pressure at the sphere surface in the dual limits $\delta \ll \ell$ and $\delta \ll R$, but for arbitrary $\ell/R$, is
	% \begin{equation}
	% n(R) = n^\infty\left( 1 + \frac{1}{6}(\ell/\delta)^2\frac{1}{1  + (\ell/R)/\sqrt{3}} \right) \, ,
	% \label{eq:nR}
	% \end{equation}
	% and
	\begin{equation}
	\Piup^{ext}(R)  = n^\infty\left(k_BT + \frac{k_sT_s}{1  + (\ell/R)/\sqrt{3}} \right) \, ,
	\label{eq:nPiR}
	\end{equation}
	while for the circle 
	\begin{equation}
	\Piup_{2D}^{ext}(R)  = n_A^\infty\left(k_BT + \frac{k_sT_s^\prime}{1  +(\ell/R) / \sqrt{2}} \right) \, ,
	\label{eq:nPiRcircle}
	\end{equation}
	where $k_sT_s^\prime = \zeta U_0^2 \tau_R/2$ is the activity in 2D.   We again see the effect of the finite run length entering as $\ell/R$
	
	%Note that in the pressure at the surface only the swim pressure is modified by the run length; the Brownian osmotic pressure is unaffected.  The Brownian displacement is small compared to the Brownian particle size---small compared to the microscopic length $\delta$---and thus  cannot be affected by the swimming activity.  
	.%~\footnote{This is only true in the dual limits $\delta \ll \ell$ and $\ell \ll R$.  For arbitrary $\delta$ and $\ell$ the full expressions must be used.}.  %For the interior problem, (\ref{eq:nchannel})-(\ref{eq:Pchannel}), the run length to channel size affects both the osmotic and swim pressures through the confinement's influence on the average number density $\langle n \rangle$; it does not affect the Brownian pressure when expressed in terms of $n_0$.
	
	For the spherical interior problem  the concentration field is given by
	\begin{equation}
	\frac{n(r)}{n(0)}    =   1 
	+   \frac{\frac{1}{6}(\ell/\delta)^2(\sinh(\lambda r)/(\lambda r) - 1)}{
		\frac{1}{6}(\ell/\delta)^2 + (1 + (\delta/R)^2)\sinh(\lambda R)/(\lambda R) - (\delta/R)^2 \cosh(\lambda R))}\, ,
	\label{eq:ninsphere}
	\end{equation}
while for the interior problem in 2D
	\begin{equation}
	\frac{n(r)}{n_A(0)} = 1+ \frac{2(\ell/\delta)^2  (I_0(\lambda^\prime r)-1)}{2(\ell/\delta)^2  + \left( 2 - (\ell/\delta)^2\right) I_0(\lambda^\prime R)+ \left( 2 + (\ell/\delta)^2\right) I_2(\lambda^\prime R)}  \, ,
	\label{eq:nincircle}
	\end{equation}
	with $I_{0,2}$ modified Bessel functions.  In the dual limits $\delta \ll \ell$ , $\delta \ll R$, the interior pressure in 2D is identical to (\ref{eq:nPiRcircle}) with $\AVE{n_A}$ replacing $n^\infty_A$.
	
	%The interior pressure in 2D is,  ,
	% \begin{equation}
	%\Pi^{int}_{2D}(R) =  \AVE{n_A}\left(k_BT + \frac{k_sT_s^\prime}{1  +(\ell/R) / \sqrt{2}}\right) \, ,
	% \label{eq:Piincircle}
	% \end{equation}
	% which is the same as the exterior pressure at the surface.
	%  \begin{equation}
	%\frac{n(r)}{n(0)} = 1 +  \frac{\frac{1}{6}(\ell/\delta)^2(\sinh(\lambda r)/(\lambda r) - 1)}{\frac{1}{6}(\ell/\delta)^2 + (1 + (\delta/R)^2)\sinh(\lambda R)/(\lambda R) - (\delta/R)^2 \cosh(\lambda R))}  \, ,
	% \label{eq:ninsphere}
	% \end{equation}
	
	% with 
	%  \begin{equation}
	% \langle n \rangle  = n_0\left( ?? \right) \, ,
	% \label{eq:aveninsphere}
	% \end{equation}
	% and 
	% \begin{equation}
	%\Pi^{int}(R) =  \AVE{n}\left(k_BT + \frac{k_sT_s}{1  + (\ell/R)/(2\sqrt{3})} \right) \, .
	% \label{eq:Piinsphere}
	% \end{equation}
	%An interesting result is that with the same active particle concentration inside and out $n^\infty = \AVE{n}$ for finite $\ell/R$, the swim pressure outside exceeds that inside, which implies that an active suspension can have a surface tension.  

	\begin{figure}%[t]
		\centering
		\includegraphics{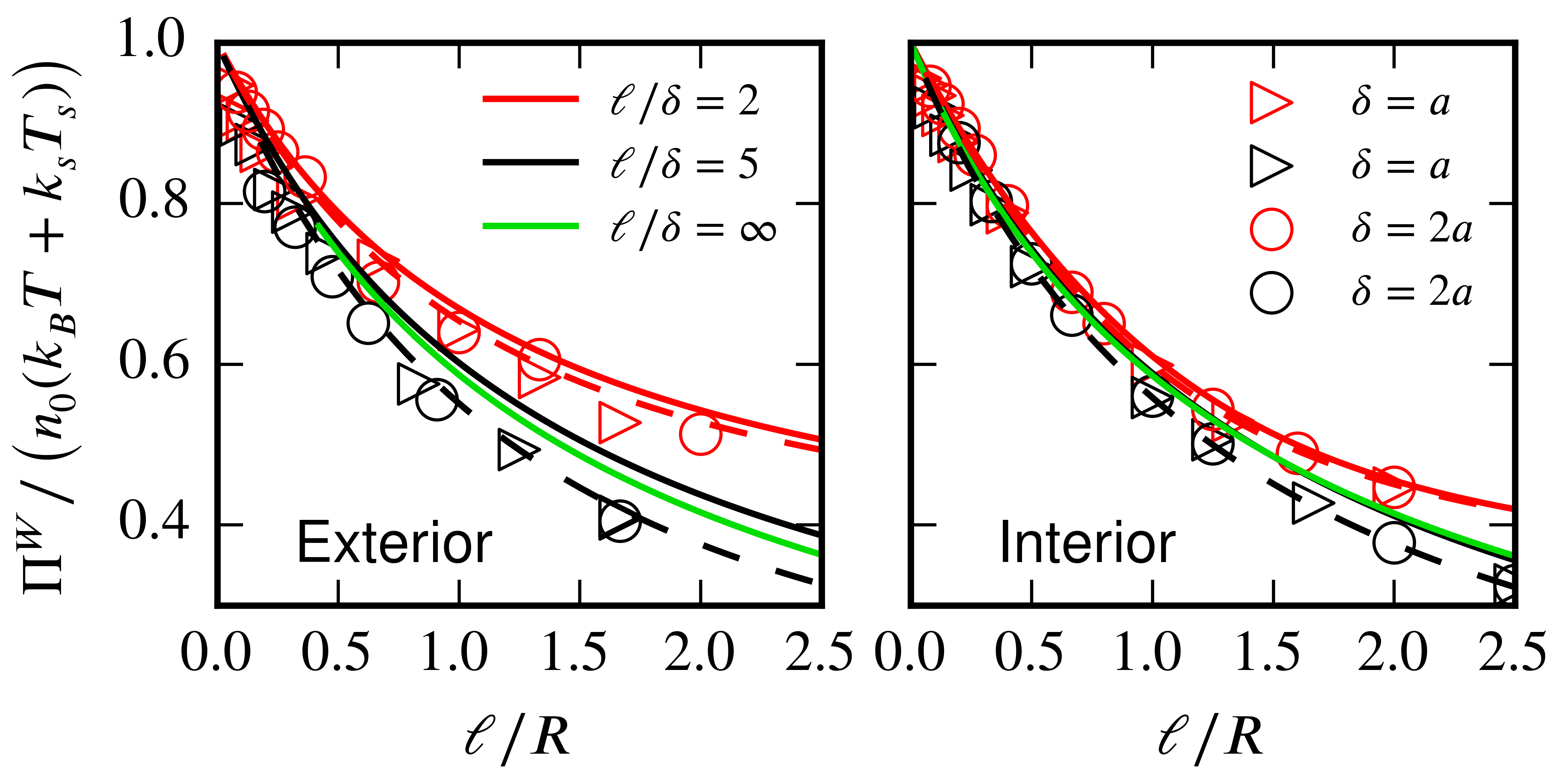}
		\caption{$\Piup^{W}$ of ABPs outside and inside a circle. Legends are as in Fig.~\ref{fig:walls}.}
		\label{fig:2Dcompare}
	\end{figure}

	Figure~\ref{fig:2Dcompare} compares the predicted results in 2D for the exterior and interior problems with ABP simulations and the next level $\Qmat$ theory. (By symmetry $\Qmat = h(r)\Imat + s(r)\bx\bx$).
	Again, the $\bbm$-level theory is quantitatively accurate unless $R/\delta < 5$.% which is not unexpected. 
	
	By symmetry there is no net force on a sphere or a circle in an active suspension.  The Brownian osmotic pressure is independent of both $\delta$ and $\ell$ (as is must be) and thus the integral of the constant Brownian osmotic pressure over the surface of {\em any} body will be zero.  
	
	From the examples   the swim pressure has a correction due to the finite run length, $\Piup^{swim} \sim k_sT_s/[1 + \alpha (\ell/R)]$, where $\alpha$ is a constant and $R$ is the curvature of the body.   Thus, in the limit $\ell/R \ll 1$ the swim pressure is a constant at each point on the body surface and there will again be no  force  no matter what its shape.   This is as we would expect from the pressure for a {\em macroscopic} object.  Only when the run length is comparable to the local radius of curvature of a body is it possible to have a net force from the swimmers' activity.  
	%Indeed, since all the action occurs within a distance of $O(\delta)$ of the body surface a boundary-layer analysis  allowing for slow variation along the surface of the body will show that the local swim pressure behaves as $\Pi^{swim} \sim k_sT_s/[1 + \alpha\,  \ell\, \nabla\cdot\bn]$, where $\nabla \cdot\bn$ is the {\em local} curvature of the body. 
	
	%That force is
	%\begin{equation}
	%\AVE{\bF}_{\bq} = -n^\infty k_sT_s \int_{S_B} \frac{\bn}{1 + \alpha\, \ell\, \nabla\cdot\bn}\,dS \, ,
	%\label{eq:netforce}
	%\end{equation}
	%with $\alpha_{3D} = 1/(2\sqrt{3})$ and $\alpha_{2D} = \sqrt{2}$.
	%

	%The general solution for the $n$- and $\bbm$-fields, which is presented elsewhere \cite{ActiveColloids2016},  can be expressed in terms of spherical harmonics (circular harmonics in 2D) times functions of the radial  coordinate.  %For an axisymmetric body $n(r,\theta) = \sum_{n=0}^\infty A_nP_n(\theta) R_n(r)$, where $P_n(\theta)$ are Legendre polynomials and 
	%The radial functions involve the both the microscopic length $\delta$ and the run length $\ell$ as seen for the sphere and circle in (\ref{eq:nsphere}) and (\ref{eq:ncircle}).  The $\bbm$-field has a similar expansion.  Application of the boundary conditions on $n$ and $\bbm$ determines the coefficients in the expansion, from which the force can be computed by integration.  

	\begin{figure}
		\centering
		\includegraphics{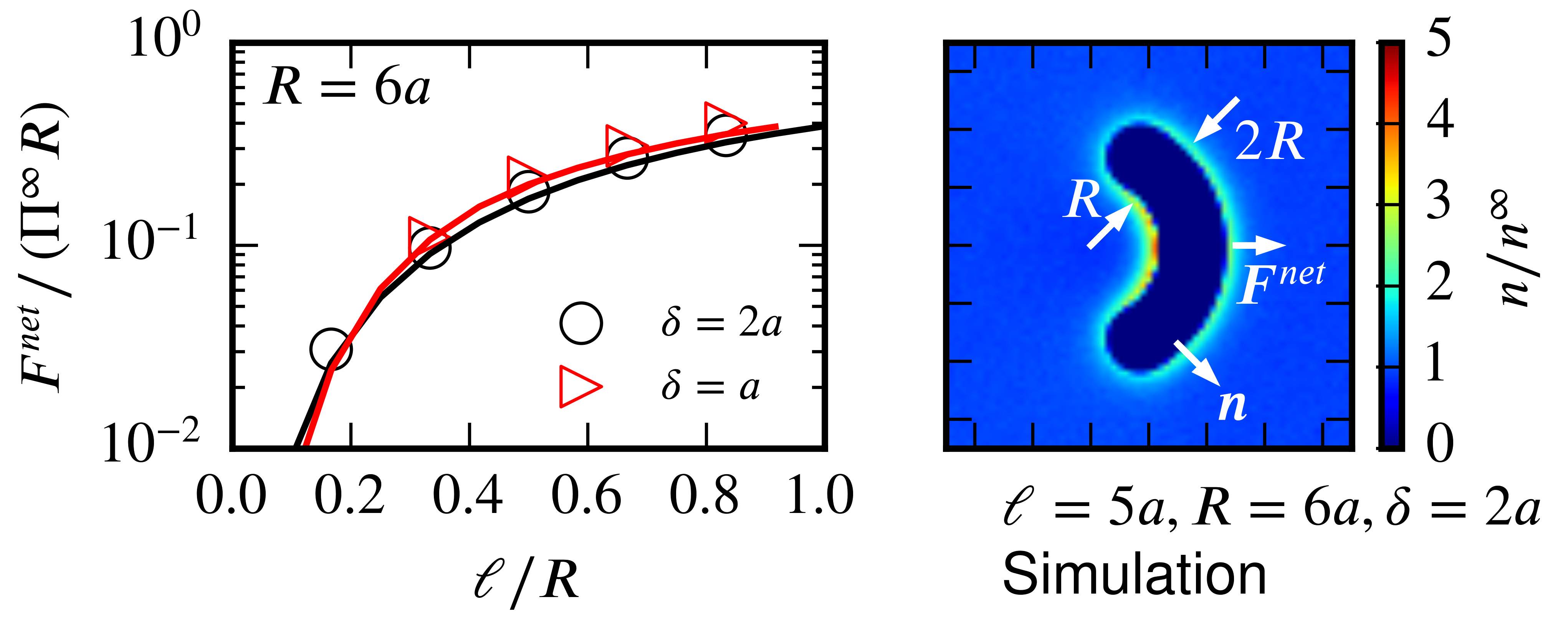}
		\caption{Theoretical predictions of the force on an asymmetric body in 2D with curvatures $R$ and $2R$ compared to ABP simulations. The symbols are simulation results and the solid lines were obtained by numerically solving for the $n,\bm{m}$ fields.  The force is calculated from~(\ref{eq:forceonbody2}) with body normal is $\bn$ and is scaled by the bulk active pressure $\Piup^\infty = n(k_BT + k_sT_s)$.}
		\label{fig:asym}
	\end{figure}

	Equation~(\ref{eq:forceonbody2}) for the force applies to any body shape and for any size body. 
	Figure~\ref{fig:asym}  compares the force on an axisymmetric body in 2D determined by solving the $n,\bbm$ fields numerically\footnote{The unsteady equations (\ref{eq:jn})-(\ref{eq:Q})  were solved numerically with a standard Galerkin P2-FEM method with adaptive mesh refinement. Implicit time-stepping was used to ensure solution stability, and the solution is tracked long enough ($\sim 150\tau_R$) to reach a steady state.} with ABP simulations.
	The agreement is excellent.   If the body were free to move, its speed will result from the balance of its Stokes drag %$-\bR_{FU}\cdot \bU$ 
	with the active force. %, where $\bR_{FU}$ is the hydrodynamic resistance tensor for the body.  
	A body may also rotate if the active pressure exerts a net torque on the body, which is given by  $\AVE{L}_{\bq} = - k_BT \oint_{S_B} (\bx - \bx_0) \times n \bn \dup S$, where $\bx_0$ is the point about which the torque is taken.  %The angular velocity then follows from the hydrodynamic resistance tensor coupling torque and rotation, $- \bR_{L\Omega}\cdot\bOmega$.

	%the force on a body in 2D composed of two circular arcs of different radii.  Ignoring the sharp corners, we can use the result (\ref{eq:nPiRcircle}) for the pressure on each arc to predict that the body should experience the force  If the body is free to move, its speed will result from the balance of its drag $-\zeta_B U$ with the active force, where $\zeta_B$ is the drag coefficient for the body.  The discrepancy between the theory and the simulation results may be due to the sharp corners where the curvature is high.
	\section{Discussion}
	From the structure of the concentration distribution and its dependence on the ratios $\ell/\delta$  and $\ell/L$ we can readily predict if a given body will experience a net force.  For example, a long thin rectangle will experience no net force or torque as the active pressures are equal on both faces.  If, however, we add a side arm to create a `T'-shaped particle, there will be a force in the direction of the top of the `T'.  To a first approximation at each point of the surface there will be a concentration boundary layer as in  (\ref{eq:n(z)})  for a flat wall and thus the active pressure will be the same at all points on the body surface.  However, where the top meets the side arm, the two solutions will superimpose giving an increased concentration in the `corners' and thus a net force (and torque if the side arm is not at the midpoint).  The precise magnitude of the force requires a detailed solution,  but that there should be a force can be simply reasoned.  Similarly, a wedge-shaped particle will experience a force towards the point of the wedge from the overlapping of the concentration boundary layers on the inside corners.  This reasoning can be continued for bodies composed of straight segments joined at angles \citep{ConfindedSwimmers2014}. %provided the screening length is small compared to a segment length.  
	The detailed magnitude of the force, of course, requires a  solution of (\ref{eq:jn})-(\ref{eq:Q}) for the given body geometry as done  in Fig.~\ref{fig:asym}.

	We can also reason about the interaction between two bodies through their disturbance to the concentration and polar-order fields.  Two bodies will experience a depletion-like attraction due to the exclusion of active particles between them.\footnote{The force can be estimated from (\ref{eq:nsphere}):  $ \bF \sim  - k_BT V \nabla n$, where $V$ is the volume of the first particle and the concentration gradient due to the second particle is evaluated at the center of the first particle.}
	When bodies are far apart the attractive force is very weak and decays exponentially with separation according to (\ref{eq:nsphere});   this exponential dependence was  seen in the simulations of \cite{Ray2014}.  Outside the screening length the  concentration is undisturbed and  the depletion interaction will be the same as for  passive colloidal particles where the exclusion zone is  geometric~\citep{Depletion1954}; the Brownian osmotic pressure on the exposed surfaces is replaced  by the active pressure that includes the run length (\ref{eq:nPiR}).  Note that the exclusion occurs on the microscopic scale $\delta$ (or swimmer size $a$), not on the scale of the run length.  Even when the gap between two particles is less than $\ell$  the active particles can still access this space and exert their swim pressure.  
	
	In the examples we have considered there was no polar order far from the boundary, nor a gradient in the concentration of swimmers, and thus force or motion can only arise if the run length is on the order of the body size, $\ell/L \sim O(1)$, {\em and} if the symmetry is broken by the body shape.  With macroscopic polar order, which can result from an orienting field applied to the swimmers \citep{PolarField2014}, even a spherical particle will experience a net force and move due to the imbalanced active pressure.  We also used the simplest no-flux boundary condition on the  polar order field at the body surface, but this condition can be modified.  For example, a portion of the body surface may be treated such that the active particles achieve a preferred orientation or experience a localized orienting field.  Such a `Janus' particle may have a net force due to a spatially varying polar order boundary condition.  
	
	Indeed,  a localized boundary orienting field was used by \cite{Solon2015b} to argue that the pressure of active matter is not a `state' function, as the force per unit area on a wall is no longer equal to the swim pressure far from the surface.  As our microscopic theory shows, this is to be expected in general:  boundary curvature, the detailed flux condition at the surface, etc. can all affect the value of the concentration at the surface and thus the force on the boundary.  We showed recently \citep{Yan2015} that the polar order induced by an orienting field acts like a body force on the active material, and when this `internal' body force is included in the momentum balance, the force per unit area on the wall plus the integral of the internal body force is equal to the active pressure far from the surface, thus restoring the active pressure as a state function.

	Furthermore, since the behavior is dominated by the rapid variations that occur on the screening length adjacent to the body surface, the situation has features in common with phoretic-like problems and hydrodynamic fluid motion can be incorporated in a manner similar to diffusiophoresis \citep{Anderson1989, Diffusio2012, Sailing2014}.
	
	The theory we have developed and applied for dilute active matter can be extended to higher concentrations of swimmers.  The $N$-particle Smoluchowski equation for passive Brownian particles including excluded volume  and full hydrodynamic interactions is well known, as is the form of the many-body hydrodynamic swim force \citep{Yan2015}.  Reduction to the lowest moments, $n$ and $\bbm$, is certain to give rise to new phenomena since the swim diffusivity, which  enters the flux expressions, can be a decreasing function of the swimmer concentration \citep{Pressure2014}.
	
	\section*{Acknowledgments}
	Discussions with S.C. Takatori and E.W. Burkholder are greatly appreciated.  This work was supported by NSF Grant No. CBET 1437570.
	
	\appendix
	\section{The Langevin equation and simulation}
	Active Brownian particles (ABP) are governed by the overdamped Langevin equation
	\begin{equation}\label{eq:ABP}
	0 = - \zeta\bU +  \bF^{swim} + \bF^{P} + \bF^B  \, ,
	\end{equation}
	where $\bU$ is the particle velocity, $\bF^P$ is a boundary force, and $\bF^B$  is the Brownian force responsible for the translational diffusivity $D_T = k_BT/\zeta$.  The orientation vector $\bq$ in the swim force is subject to rotational Brownian diffusion ($D_R = 1/\tau_R$), and follows directly from a torque  balance. For a spherical swimmer, $\zeta=6\upi\eta a$, where $a$ is the swimmer radius and $\eta$ is  the viscosity of the suspending Newtonian fluid.  For Brownian reorientation, $D_R = k_BT/8\upi \eta a^3$, and the microscopic length scale $\delta = \sqrt{D_T/D_R} = \sqrt{4/3}a$.
	
	In the Brownian dynamics simulations each swimmer is a sphere of radius $a$ and the Langevin equation  is integrated to track the  dynamics. When a swimmer hits a boundary, it experiences a hard-particle force $\bF^P$ to prevent it from penetrating the boundary.  (Following \citep{Foss2000} a potential-free hard particle force is implemented.)
 By Newton's Third Law the boundary experiences an opposite force $-\bF^P$, and then $\Piup^W$ is calculated from the definition of the pressure: $\sum\bF^P/A$ in 3D or $\sum\bF^P/L$ in 2D, where $A$ and $L$ are the wall area and length, respectively.

	Swimmer-swimmer collisions are ignored because we compare simulation with the dilute theory. Each simulation is tracked long enough to ensure that good steady state statistics are achieved. In some cases the simulation time is as long as $3000\tau_R$.  
	
	\section{Closure of the Smoluchowski equation}
%	The Smoluchowski equation for the probability density for finding a swimmer at $\bx$ with orientation $\bq$ is
%	\begin{equation}
%	\frac{\partial P(\bx,\bq,t)}{\partial t} + \nabla \cdot \bj^T + \nabla_{R} \cdot \bj^R = 0 \, ,
%	\label{eq:SmolappB}
%	\end{equation}
%	where the translational and rotational  fluxes are
%	\begin{eqnarray}
%	\bj^T & =&   (U_0 \bq - D_T \nabla \ln P)P\, , \\*
%	\bj^R & = &  - D_R \nabla_{R} P\, ,
%	\label{eq:fluxes}
%	\end{eqnarray}
%	and $\nabla_R = \bq\times \nabla_{\bq}$ is  the orientational gradient operator.

	As is standard \citep{Saintillan2015}, the first few moments of the Smoluchowski equation (\ref{eq:Smol}) are:
	\begin{eqnarray}
	\frac{\partial n}{\partial t}  +   \nabla\cdot \bj_n = 0\, , & \quad  & \bj_n   =  U_0\bbm - D_T \nabla n\, , \label{eq:n}\\*
	\frac{\partial \bbm}{\partial t}  +   \nabla\cdot  \bj_m + 2D_R \bbm = 0\, , & \quad & \bj_m  =   U_0\tilde{\Qmat}  - D_T \nabla\bbm\, , \label{eq:m} \\*
	\frac{\partial \tilde{\Qmat}}{\partial t}  +   \nabla\cdot  \bj_{\tilde{\mathsfi{Q}}} + 6D_R[\tilde{\Qmat} - \textstyle{\frac{1}{3}} n \Imat] = 0\, ,&  \quad & \bj_{\tilde{\mathsfi{Q}}} =  U_0\tilde{\Bmat}  - D_T \nabla\tilde{\Qmat} \, .
	\label{eq:QappB}
	\end{eqnarray}
%	with  fluxes 
%	\begin{eqnarray}
%	\bj_n & =&  U_0\bbm - D_T \nabla n \, , \\*
%	\bj_m & =&   U_0\tilde{\Qmat}  - D_T \nabla\bbm \, , \\*
%	\bj_{\tilde{\mathsfi{Q}}}& =&  U_0\tilde{\Bmat}  - D_T \nabla\tilde{\Qmat} \, .
%	\end{eqnarray}
	Here, we have written the second moment as $\tilde{\Qmat}(\bx,t) = \int \bq\bq P(\bx,\bq,t)\mathrm{d}\bq$, and $\tilde{\Bmat} = \int \bq\bq\bq P \mathrm{d}\bq$ is the third moment.
	
	In the simplest situation of no temporal or spatial variation,  a uniform concentration $n$ and no polar order $\bbm = 0$ are solutions of (\ref{eq:n})-(\ref{eq:m}), and the second moment has solution $\tilde{\Qmat} = \textstyle{\frac{1}{3}} n\, \Imat$.  This leads to the natural definition of the nematic order field 
	%\begin{equation}
	$\tilde{\Qmat} = \Qmat + \textstyle{\frac{1}{3}} n \Imat$ , %\, \Imat
	%\end{equation}
	or
	%\begin{equation}
	$\Qmat(\bx,t) = \int (\bq\bq- \textstyle{\frac{1}{3}}\Imat) P(\bx,\bq,t)\mathrm{d}\bq$.
	%\end{equation}
	The conservation equation for $\Qmat$ is 
	\begin{equation}
	\frac{\partial \Qmat}{\partial t}  +   \nabla\cdot  \bj_{\mathsfi{Q}} + 6D_R\Qmat  = 0\, ,
	\label{eq:QQ}
	\end{equation}
	which now does have the solution of no nematic order $\Qmat = 0$ in the uniform case.  The 
	flux expressions now become: $\bj_n =  U_0\bbm - D_T \nabla n$, $\bj_m  =   U_0\Qmat  +  \textstyle{\frac{1}{3}}U_0 n\, \Imat - D_T \nabla\bbm$ and $\bj_{Q}  =  U_0\tilde{\Bmat}  - \textstyle{\frac{1}{3}} U_0   \bbm \, \Imat - D_T \nabla\Qmat$. 
	
%	\begin{eqnarray}
%	\bj_n & =&  U_0\bbm - D_T \nabla n \, , \\*
%	\bj_m & =&   U_0\Qmat  +  \textstyle{\frac{1}{3}}U_0 n\, \Imat - D_T \nabla\bbm \, , \\*
%	\bj_{Q} & =&  U_0\tilde{\Bmat}  - \textstyle{\frac{1}{3}} U_0   \bbm \, \Imat - D_T \nabla\Qmat \, .
%	\end{eqnarray}

	We shall discuss the $\tilde{\Bmat}$-field and its closure in a moment, but we can already appreciate why closing the equations with $\Qmat = 0$ leads to a very good approximation as demonstrated by the results presented in the main text.  First, we are not setting the second moment to zero; we are approximating the second moment with the `isotropic' distribution $\tilde{\Qmat} \approx \textstyle{\frac{1}{3}}   n\, \Imat $.  Second, 
	(\ref{eq:QQ}) shows that the $\Qmat$-field is screened like the $\bbm$-field, but with a temporal decay that is 3-times faster and a screening length that is $\sqrt{3}$ shorter.  Third, as we show below, when variations are slow, like the $\bbm$-field where $\bbm \sim - \frac{1}{6} (U_0/D_R) \nabla n$, the nematic order goes as $\Qmat \sim - (U_0/D_R) \nabla \bbm \sim  (U_0/D_R)^2\nabla\nabla n$, % \sim (\ell/L)^2 \bar{\nabla} \bar{\nabla} n$, where $\nabla$ has been scaled on the macroscopic length scale $L$.  
	and thus $\Qmat \sim O(\ell/L)^2n$  which is small.  Finally, for the 1D flat wall problem, the value of the concentration at the surface, $n^\infty(1 + \frac{1}{6}(\ell/\delta)^2)$, follows directly from the full Smoluchowski equation and is {\em independent} of the closure.  Thus, it is perhaps not surprising that the simple closure $\Qmat = 0$ works very except when the body curvature is on the order of the microscopic length $\delta = \sqrt{D_T \tau_R}$.
	
	The equation for the third moment is 
	\begin{equation}
	\frac{\partial \tilde{\Bmat}}{\partial t}  +   \nabla\cdot  \bj_{\tilde{\mathsfi{B}}} + 12D_R[\tilde{\Bmat} - \textstyle{\frac{1}{6}}{\balpha} \cdot\bbm]  = 0\, , \quad \bj_{\tilde{\mathsfi{B}}} = U_0\tilde{\Cmat} - D_T\nabla \tilde{\Bmat} \, ,
	\end{equation}
	where $\alpha_{ijkl} = \delta_{ij}\delta_{kl} + \delta_{ik}\delta_{jl} + \delta_{il}\delta_{jk}$ is the fourth order isotropic tensor and $\tilde{\Cmat} = \int \bq\bq\bq\bq P(\bx, \bq)\mathrm{d}\bq$ is the fourth moment.  
	%The $\tilde{\Bmat}$ flux is
	%\begin{equation}
%	\bj_{\tilde{\mathsfi{B}}} = U_0\tilde{\Cmat} - D_T\nabla \tilde{\Bmat} \, ,
%	\end{equation}
%	where $\tilde{\Cmat} = \int \bq\bq\bq\bq P(\bx, \bq)\mathrm{d}\bq$ is the fourth moment.
	
	The proper `isotropic' $\tilde{\Bmat}$ field is $\tilde{\Bmat} = \Bmat + \textstyle{\frac{1}{5}} {\balpha} \cdot \bbm$, 
%	\begin{equation}
%	\tilde{\Bmat} = \Bmat + \textstyle{\frac{1}{5}} {\balpha} \cdot \bbm \, ,
%	\label{eq:BB}
%	\end{equation}
	and the equation for $\Bmat$ becomes
	\begin{equation}
	\frac{\partial \Bmat}{\partial t}  +   \nabla\cdot  \bj_{\mathsfi{B}} + 12D_R\Bmat = 0\, , \quad \bj_{\mathsfi{B}} = U_0\tilde{\Cmat} - \textstyle{\frac{1}{5}}U_0 \balpha\cdot[\Qmat + \textstyle{\frac{1}{3}}n \Imat] - D_T\nabla \Bmat \, .
	\end{equation}
%	with flux
%	\begin{equation}
%	\bj_{\mathsfi{B}} = U_0\tilde{\Cmat} - \textstyle{\frac{1}{5}}U_0 \balpha\cdot[\Qmat + \textstyle{\frac{1}{3}}n \Imat] - D_T\nabla \Bmat \, .
%	\end{equation}
		In the examples where we included the nematic field $\Qmat$, we closed the equations by setting $\Bmat = 0$, which follows the same reasons as for setting $\Qmat = 0$.  With this closure the $\Qmat$-field flux is
	\begin{equation}
	\bj_{\mathsfi{Q}} =  \textstyle{\frac{1}{5}}U_0[\balpha    - \textstyle{\frac{5}{3}}   \Imat\Imat] \cdot \bbm- D_T \nabla\Qmat \, ,
	\end{equation}
	which was used in the examples presented in the main text.  With this constitutive equation for 
	the flux, for slow variations we see that $\Qmat \sim (\ell^2/135)(\nabla \nabla - \frac{1}{3}\Imat\nabla^2)n$.

%	We have considered the simplest situation in which there is no net polar order far from a boundary and thus (\ref{eq:m}) governs  $\bbm$.  With an external field that tends to align the swimmers and biases their motion, for example, an external torque or hydrodynamic shearing flows, the conservation equation  for $\bbm$ now has an additional `sink' term, which can be written as $2D_R[\bbm - \bbm^\infty]$ where $\bbm^\infty$ is the polar order  far from the boundary induced by the field.  The equation for $\Qmat$ will have a similar $\Qmat^\infty$ term.  One now writes conservation equations for the departures of the polar order and nematic fields from their undisturbed values, $\bbm^\prime = \bbm - \bbm^\infty$ and $\Qmat^\prime = \Qmat - \Qmat^\infty$, and then closes the disturbance equations along the lines done here.   It is not yet known if this simple closure proves as accurate when there is net bulk polar order.

%	For the spherical interior problem in 3D the concentration field is given by
%	\begin{equation}
%	\frac{n(r)}{n(0)}    =   1 
%	+   \frac{\frac{1}{6}(\ell/\delta)^2(\sinh(\lambda r)/(\lambda r) - 1)}{
%		\frac{1}{6}(\ell/\delta)^2 + (1 + (\delta/R)^2)\sinh(\lambda R)/(\lambda R) - (\delta/R)^2 \cosh(\lambda R))}\, .
%	\label{eq:ninsphere}
%	\end{equation}

\bibliographystyle{jfm}
	\bibliography{FOB}

\end{document}